# Decisive Influence of Cation Size on the Magnetic Groundstate and Non-Fermi Liquid Behavior of $ARuO_3$ (A = Ca, Sr)

G. Cao, O. Korneta, S. Chikara and L. E. DeLong

Department of Physics and Astronomy, University of Kentucky, Lexington, KY 40506

P. Schlottmann

Physics Department, Florida State University, Tallahassee, FL 32306

We report calorimetric, magnetic and electric transport properties of single-crystal $CaRuO_3$ and $SrRuO_3$ as a function of temperature $T$ and applied magnetic field $B$. We find that $CaRuO_3$ is a non-Fermi-liquid metal near a magnetic instability, as characterized by the following properties: (1) the heat capacity $C(T,B) \sim -T \log T$ is readily enhanced in low applied fields, and exhibits a Schottky peak at 2.3 K that exhibits field dependence when $T$ is reduced; (2) the magnetic susceptibility diverges as $T^{\gamma}$ at low temperatures with $1/2 < \gamma < 1$, depending on the applied field; and (3) the electrical resistivity exhibits a $T^{3/2}$ dependence over the range $1.7 < T < 24$ K. No Shubnikov-de Haas oscillations are discerned at $T = 0.65$ K for applied fields up to 45 T. These properties, which sharply contrast those of the itinerant ferromagnet $SrRuO_3$, indicate $CaRuO_3$ is a rare example of a stoichiometric oxide compound that exhibits non-Fermi-liquid behavior near a quantum critical point.

A primary characteristic of the 4d-transition elements is that their 4d-orbitals are more extended than those of their 3d-electron counterparts. Stronger p-d hybridization and electron-lattice coupling, along with reduced intra-atomic Coulomb interaction U, are thus anticipated in these systems. Consequently, the 4d-transition metal oxides have comparable U and bandwidth W, and are precariously balanced on the border between metallic and insulating behavior and/or on the verge of long-range magnetic order. Consequently, small perturbations such as slight changes in lattice parameters, application of magnetic field, etc., can readily tip the balance, inducing drastic changes in the ground state, such as the inducement of long-range magnetic order in $CaRuO_3$ by merely a few percent of impurity doping [1-5].

Isostructural and isoelectronic $CaRuO_3$ and $SrRuO_3$ are "n = ∞" members (n = number of Ru-O layers/unit cell) of the Ruddlesden-Popper (RP) series. $Ca_{n+1}Ru_nO_{3n+1}$ and $Sr_{n+1}Ru_nO_{3n+1}$, respectively, have been extensively studied [1-14], and their sharp differences in magnetic behavior are classic examples of the sensitivity of 4d-electron band structure to structural distortions. Both compounds are orthorhombic, but $SrRuO_3$ has a more "ideal" and less distorted perovskite structure and is an itinerant ferromagnet with a Curie temperature $T_C = 165$ K and a saturation moment $M_s$ of 1.10 $\mu_B$/Ru with an easy axis in the basal plane [2]. The relatively large extension of the 4d-orbitals leads to a CEF splitting of the ($Ru^{4+}$) $4d^4$ configuration that is large enough for Hund's rules to partially break down, yielding a low-spin state with S = 1 ($^3T_{1g}$). On the other hand, the isostructural and isoelectronic $CaRuO_3$ forms in the same crystal structure as $SrRuO_3$, but with a rotation of $RuO_6$ octahedra which is approximately twice as large as that observed for $SrRuO_3$, due to ionic size mismatches between Ca and Ru ions (ionic radius $r$ = 1.00

Å and 1.18 Å for $Ca^{2+}$ and $Sr^{2+}$, respectively, vs. r = 0.620 Å for $Ru^{4+}$). This makes $CaRuO_3$ less favorable for ferromagnetism due to a weaker exchange interaction *U*; thus, $Ug(E_F) < 1$, where $g(E_F)$ is the density of states at the Fermi energy [7].

More generally, the physical properties of the entire RP series, $Ca_{n+1}Ru_nO_{3n+1}$ and $Sr_{n+1}Ru_nO_{3n+1}$, critically depend on the deformations and relative orientations of the corner-shared RuO octahedra that, in turn, determine the crystalline field splitting and the band structure. Consequently, inter- and intra-layer magnetic couplings and the ground state are critically linked to n and to the cation (Ca or Sr): the $Sr_{n+1}Ru_nO_{3n+1}$ compounds are metallic and tend to be ferromagnetic (with $Sr_2RuO_4$ [15] being an exception), whereas the $Ca_{n+1}Ru_nO_{3n+1}$ compounds are all at the verge of a metal-nonmetal transition and prone to antiferromagnetism.

The trend for the magnetic ordering temperature with the number of directly coupled Ru-O layers n is also surprisingly different for these two isostructural and isoelectronic systems. The Curie temperature $T_c$ increases with n for $Sr_{n+1}Ru_nO_{3n+1}$; whereas, the Neel temperature, $T_N$, decreases with n for $Ca_{n+1}Ru_nO_{3n+1}$, as shown in **Fig. 1**. From the properties of the compounds we assigned a sequence of approximate *W/U* ratios, so that the two RP series can be placed into one phase diagram. Such a drastic dependence of the ground state on cation has not been observed in other transition metal RP systems, which implies the critical roles of the lattice and orbital degrees of freedom in the properties of these materials. We note that $CaRuO_3$ is located near a border that separates ferromagnetic and antiferromagnetic groundstates, which is a situation that almost always leads to extraordinary properties.

Indeed, there are several other examples of intriguing quantum phenomena occurring in itinerant-electron materials that are on the borderline; for example, p-wave superconductivity in $Sr_2RuO_4$ [15], superconductivity and ferromagnetism in $ZrZn_2$ [16, 17] and URhGe [18], a ferromagnetic quantum critical point (QCP) in MnSi under pressure [19], a metamagnetic transition with QCP end-point tuned by a magnetic field in $Sr_3Ru_2O_7$ [20], a QCP with anomalous ferromagnetism in $Sr_4Ru_3O_{10}$ [21] and, most recently, a QCP with ferromagnetic fluctuations in $SrIrO_3$ [22].

In this paper, we report phenomena associated with a possible QCP and strong spin fluctuations leading to the breakdown of Fermi-liquid behavior in $CaRuO_3$, including a divergent specific heat [$C/T \sim -\log T$] followed by a Schottky anomaly with decreasing $T$, and unusual power laws in resistivity $\rho$ and magnetic susceptibility $\chi$ at low temperatures. Surprisingly, no Shubnikov-de Haas effect is discerned in high-quality single crystals of $CaRuO_3$ at temperatures as low as 0.65 K and in high applied magnetic fields up to 45 T. Stoichiometric $CaRuO_3$ can be readily tuned with very modest magnetic fields, and this makes it an interesting model system for experimental and theoretical studies. These underlying properties sharply contrast those of the isostructural and isoelectronic $SrRuO_3$, underscoring the decisive influence of lattice degree of freedom on the ground state in the ruthenates.

Single crystals of $CaRuO_3$ and $SrRuO_3$ were grown using flux techniques. All Single crystals were grown in Pt crucibles from off-stoichiometric quantities of $RuO_2$, $CaCO_3$ ($SrCO_3$) and $CaCl_2$ ($SrCl_2$) mixtures with $CaCl_2$ ($SrCl_2$) being a self flux. They were characterized by single crystal or powder x-ray diffraction, EDS and TEM. No impurities or intergrowths were found. Heat capacity measurements were performed with a Quantum Design PPMS that utilizes a thermal-relaxation calorimeter operating in fields up to 9 T. Magnetic and transport properties below 7 T were measured using a Quantum Design MPMS 7T LX SQUID Magnetometer equipped with a Linear Research Model 700 AC bridge. The ultra-high magnetic field measurements were conducted at National High Magnetic Field Laboratory in Tallahassee, Florida.

**Fig. 2a** shows the DC magnetic susceptibility $\chi$ as a function of $T$ at $B = 0.1$ T for **B** || **c**-axis ($\chi_c$) and **B** $\perp$ **c**-axis ($\chi_{ab}$). Both $\chi_c$ and $\chi_{ab}$ are strongly temperature-dependent, reaching 9 x $10^{-3}$ emu/mole at $T$ = 1.7 K. For comparison, $\chi_{ab}$ at B = 0.01 T for isoelectronic $SrRuO_3$ is also shown in **Fig. 2a** (right scale), showing a sharp transition at the Curie temperature $T_c$ = 165 K. The reciprocal susceptibilities $\chi_c^{-1}$ and $\chi_{ab}^{-1}$ for $CaRuO_3$ display linear $T$-dependences, consistent with a Curie-Weiss behavior for $T >$ 100 K. A Curie-Weiss fit for 140 <T <320 K yields an effective moment of 2.2 $\mu_B$/Ru and a Curie-Weiss temperature, $\theta_{cw}$, of -218 K. This would normally suggest antiferromagnetic interactions among the Ru spins, and is consistent with the reduction of the susceptibility in a magnetic field (**Fig. 2b**). However, the large value of $\theta_{cw}$ (compared to measurement temperature) makes this explanation dubious. For $1.7 < T <$ 25 K, $\chi_{ab}^{-1}$ (and $\chi_c^{-1}$, not shown) follows non-standard power laws that range from $T^{1/2}$ for $B < 1.5$ T to $T^{\alpha}$ with $\alpha > 1$ for $B > 1.5$ T, as shown in **Fig. 2b**. Note the brief deviation from the linear dependence of $\chi_{ab}^{-1}$ vs. $T^{1/2}$ below 2.5 K (indicated by an arrow). Nevertheless, the high sensitivity of the temperature exponent to low applied magnetic fields suggests a proximity to a magnetic instability. Similar behavior is also seen in the ferromagnetic QCP system $SrIrO_3$ [22].

The low temperature specific heat $C(T, B)$ data acquired over $1.8 < T < 25$ K and $B < 9$ T offer important insights into the low energy excitations of $CaRuO_3$. For $T > 18$ K, the specific heat is well described by $C(T) = \gamma T + \beta T^3$ with $\gamma = 38.5$ mJ/mole $K^2$ and $\beta = 0.10$ mJ/mole $K^4$, yielding a Debye temperature of 455 K, and suggesting that only electronic and phonon contributions are significant in this temperature range (data not shown). In this range, C is also seen to be insensitive to B, which is an indication that the thermal energy is much larger than the magnetic energy in this temperature interval. The sizable $\gamma$-value implies that renormalizations of the effective mass are significant above 18 K.

The distinct temperature dependence for the two isoelectronic compounds is clearly illustrated in **Fig. 3a**, which shows $C/T$ vs $T$ for $CaRuO_3$ and $SrRuO_3$ for $1.8 < T < 30$ K at $B = 0$. Extrapolation to $T = 0$ yields $C/T$ (or $\gamma$) to be 71 and 27 mJ/mole-$K^2$ for $CaRuO_3$ and $SrRuO_3$, respectively, suggesting that the density of states for the former is clearly larger than that of the latter. Furthermore, $C(T)$ for $SrRuO_3$ is predominantly proportional to $T^{3/2}$ throughout the entire range of $1.8 < T < 25$ K (see right and upper axis as indicated by the arrows in **Fig. 3a**) as expected for magnon excitations out of a ferromagnetically ordered state. In sharp contrast, $C(T)$ for $CaRuO_3$ is proportional to $-T \log(T)$ below 13 K, which is a signature of the breakdown of Fermi-liquid behavior, such as could occur for a vanishing Fermi temperature ($T_F \rightarrow 0$). Interestingly, this behavior is accompanied by a broad peak seen near 2.3 K at $B = 0$, which becomes more prominent with increasing $B$ and shifts to 2.5 K at 9 T, as shown in **Fig. 3b**. Since $\chi(T)$ shows no corresponding transition, it is likely that this peak represents a Schottky anomaly generated by the Zeeman splitting of the Ru spin degrees of freedom. With increasing $B$, as the peak shifts to higher temperatures, the anomaly also broadens (see **Fig. 3b**) transferring entropy from low temperatures to higher temperatures.

The amplitude of the logarithmic term grows slightly with increasing field up to $B$ = 1.5 T, above which it weakens, but never completely disappears. It is noted that $C/T$ at 9 T still exhibits the logarithmic term for $6 < T < 13$ K, which is bordered by a flat region for $2.5 < T < 6$ K, signaling the recovery of the Fermi-liquid behavior in high fields and low temperatures.

The detailed field dependence of C/$T$ reveals a peak at $B$ = 1.5 T that separates a regime for $B$ < 1.5 T where C/$T$ ~ *-log (T)* increases with $B$, from the complementary regime for $B$ > 1.5 T where the log ($T$) dependence weakens. The peak fades and C/$T$ becomes much less field dependent for $T$ > 4 K. This behavior is also seen in $SrIrO_3$ where the proximity to a QCP has been established [22]. It is also worth mentioning that the Wilson ratio, $R_W \equiv 3\pi^2 k_B^2\chi/\mu_B^2\gamma$, is 7.8 at $T$ = 1.8 K and shows field dependence, as shown in **Fig. 3c**. It is noticeably larger than the values (e.g., $R_W$ ~1-5) typical of heavy Fermi liquids and exchange-enhanced paramagnets such as Pd [23]. $R_W$ drops with increasing $B$, but remains substantial (7.2) at 5 T.

The presence of the quantum critical fluctuations in $CaRuO_3$ is further corroborated by the temperature dependence of the **a**-axis resistivity, $\rho_a$, as a function of $T$, shown in **Fig. 4a**. The residual resistivty $\rho_o$ is 0.013 mΩ cm and the residual resistance ratio RRR = 14. An interesting feature is that $\rho_a$ exhibits a $T^{3/2}$ law over a temperature range of $1.7 < T < 24$ K, as shown in **Fig. 4a**, where $\rho_a$ vs $T^2$ (upper scale) is also shown for comparison. **Fig. 4b** sharply contrasts **Fig. 4a**, which shows the temperature dependence in the same temperature range for $SrRuO_3$ ($\rho_o$ = 2 μΩ cm and RRR =

80); $SrRuO_3$ clearly exhibits the $T^2$-dependence (upper scale), expected for the Fermi liquid; **Fig. 4b** also shows $\rho_a$ vs $T^{3/2}$ for comparison. The $T^{3/2}$ power-law is thought to be associated with effects of diffusive electron motion caused by strong interactions between itinerant electrons and critically damped, very-long-wavelength magnons [16] and is observed in QCP systems such as MnSi [19], $Sr_3Ru_2O_7$ [20], $Sr_4Ru_3O_{10}$ [21], and $SrIrO_3$ [22].

**Fig. 4c** displays $\rho_a$ as a function of magnetic field $B$ up to 45 T at $T$=0.65 K for a few representative orientations. As seen, while $\rho_a$ rises by more than one order of magnitude from 11 T to 45 T, no anomaly is observed and our data analysis using a Fast Fourier Transformation (FFT) reveals no oscillatory behavior. Such measurements were performed on several high-quality single crystals of $CaRuO_3$. Since the quasi-particles close to the Fermi surface are not well-defined near a QCP, it is not entirely surprising that the Shubnikov-de Haas effect, a manifestation of the quantization of quasi-particle orbits in a magnetic field, is not detected in $CaRuO_3$ single crystals. Another possibility for the absence of oscillations is that there are no closed orbits in the band structure, but this contradicts the present experimental and theoretical picture for $SrRuO_3$ [24, 25]. Nevertheless, the absence of the quantum oscillations in $CaRuO_3$ is particularly intriguing given the fact that the Shubnikov-de Haas effect is a fairly common occurrence in other layered ruthenates, including $SrRuO_3$ [24], $Sr_4Ru_3O_{10}$ [26], $Ca_3Ru_2O_7$ [27, 28], and $BaRuO_3$ [29].

Given the observed physical properties appear to be dominated by strong spin fluctuations, it is compelling to ascribe the non-Fermi-liquid behavior in $CaRuO_3$ to a proximity to a QCP. The single-crystal $CaRuO_3$ sample is very likely to be a *stoichiometric oxide*; and given its unusual sensitivity to low magnetic fields, it makes an outstanding model system for studies of quantum criticality in the ruthenates.

**Acknowledgement**: This work was supported by NSF through grants DMR-0240813 and 0552267. L.E.D. is supported by U.S. DoE Grant DE-FG02-97ER45653. P.S. is supported by the DOE through grant DE-FG02-98ER45707.

**Captions:**

**(Color online) Fig.1**. Phase diagram (temperature T vs bandwidth W) qualitatively describing (Ca, Sr)$_{n+1}$Ru$_n$O$_{3n+1}$. The phase diagram is generated based on our own studies on single crystals of (Ca,Sr)$_{n+1}$Ru$_n$O$_{3n+1}$ except for $Sr_2RuO_4$ and $Sr_3Ru_2O_7$. As illustrated, the ground state can be readily changed by changing the cation, and physical properties can be systematically tuned by altering the number of Ru-O layers, n. The crossover defines a possible quantum critical region where $CaRuO_3$ is situated. SC stands for superconductor, FM-M ferromagnetic metal, AFM-I antiferromagnetic insulator, PM-M paramagnetic metal, M-M magnetic metal.

**(Color online) Fig.2.** (a) The magnetic susceptibility $\chi$ as a function of temperature at B= 0.1 T for B || **c**-axis ($\chi_c$) and B⊥**c**-axis ($\chi_{ab}$). $\chi$ for $SrRuO_3$ is also shown for comparison (right scale); (b) Reciprocal susceptibility $\chi_{ab}^{-1}$ for $CaRuO_3$ as a function of $T^{1/2}$ and T (right and upper scales, for B ≥ 1.5T marked by an arrow). The behavior of $\chi_c^{-1}$ is similar and not shown.

**(Color online) Fig.3.** (a) The specific heat C divided by temperature, C/T, vs. log T for $CaRuO_3$ and $SrRuO_3$; for the latter at B=0 and for 1.8<T<24 K; C vs. $T^{3/2}$ (right and upper scales) is also shown; (b) C/T vs. log T for $CaRuO_3$ for B=0, 0.5, 1.2, 5 and 9T, and for 1.8<T<21 K. (c) C/T vs.B for some representative temperatures; the Wilson ratio $R_w$ (right scale) at T=1.8 K is also shown.

**(Color online) Fig.4.** The **a**-axis resistivity, $\rho_a$ as a function of $T^{3/2}$ and $T^2$ (right and upper scales) (a) for $CaRuO_3$ and (b) for $SrRuO_3$ for comparison; (c) $\rho_a$ $CaRuO_3$ for as a function of B up to 45 T at T=0.65 K for a few representative orientations.

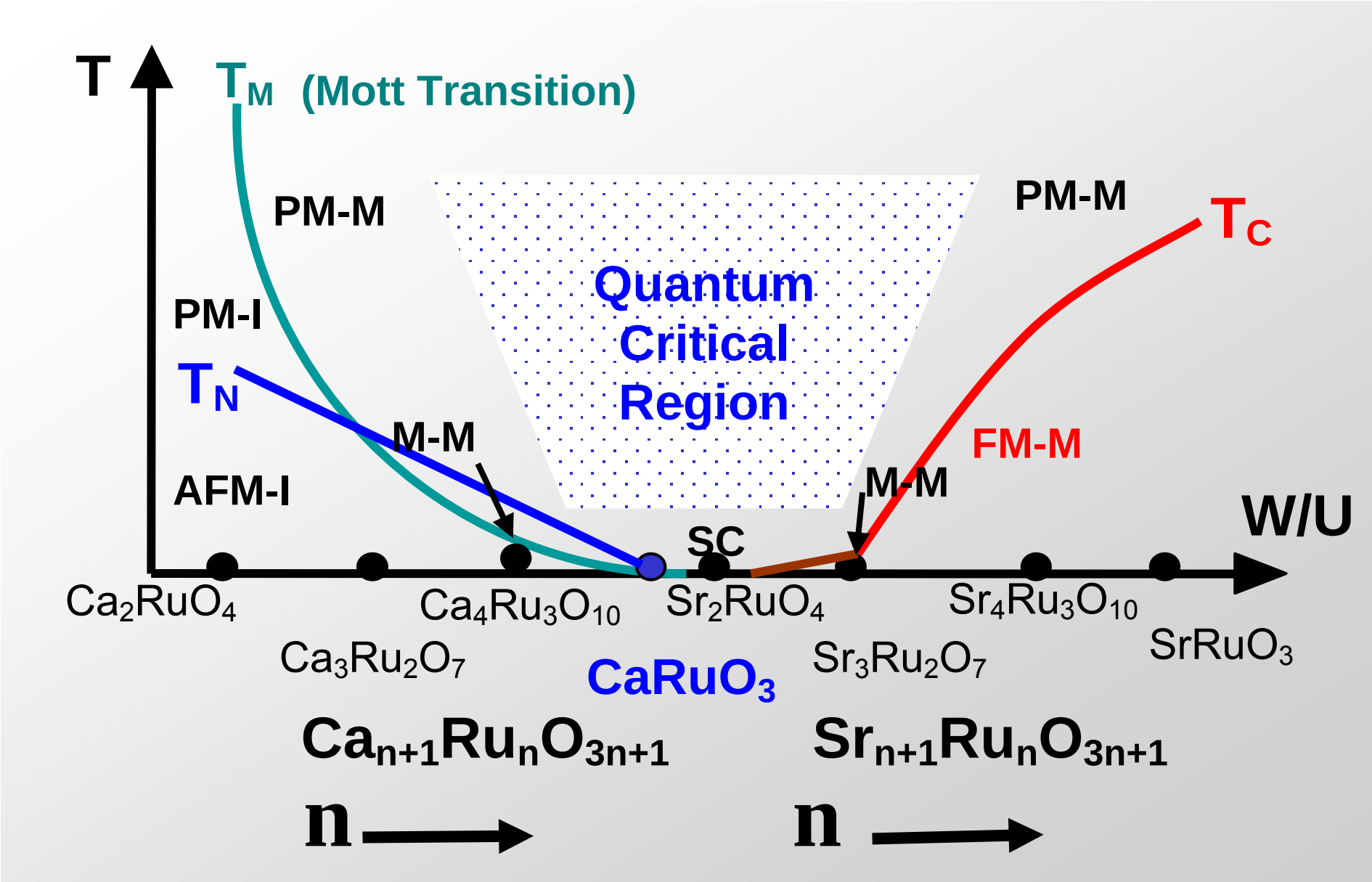

-

Fig.1

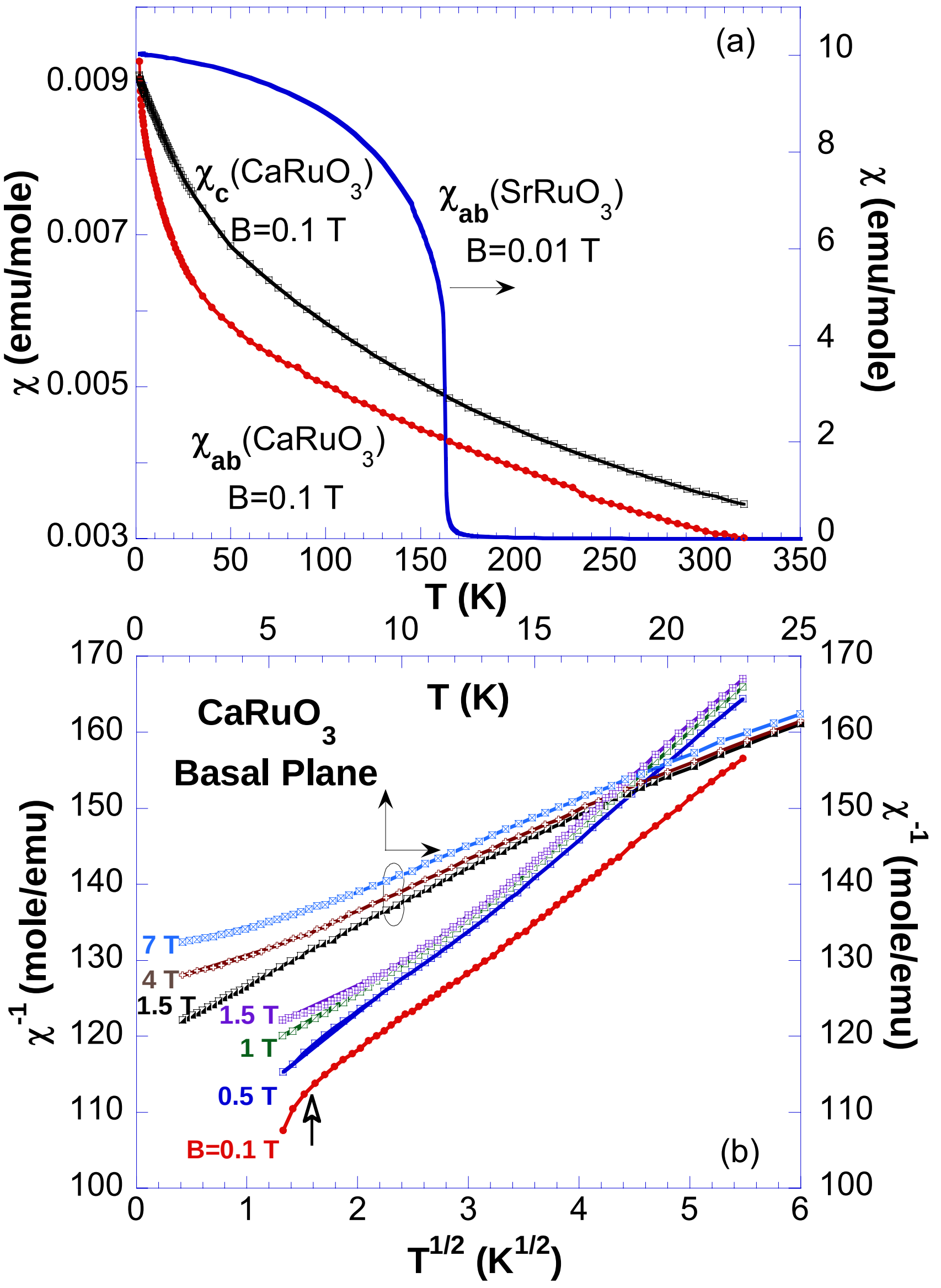
(a)
0.009
0.007
0.005
0.003
χ (emu/mole)
$\chi_c$($CaRuO_3$)
B=0.1 T
$\chi_{ab}$($SrRuO_3$)
B=0.01 T
$\chi_{ab}$($CaRuO_3$)
B=0.1 T
10
8
6
4
2
0
χ (emu/mole)
0 50 100 150 200 250 300 350
T (K)
0 5 10 15 20 25
T (K)
$CaRuO_3$
Basal Plane
$\chi^{-1}$ (mole/emu)
170 160 150 140 130 120 110 100
7 T
4 T
1.5 T
1.5 T
1 T
0.5 T
B=0.1 T
(b)
$\chi^{-1}$ (mole/emu)
0 1 2 3 4 5 6
$T^{1/2}$ ($K^{1/2}$)

Fig.2

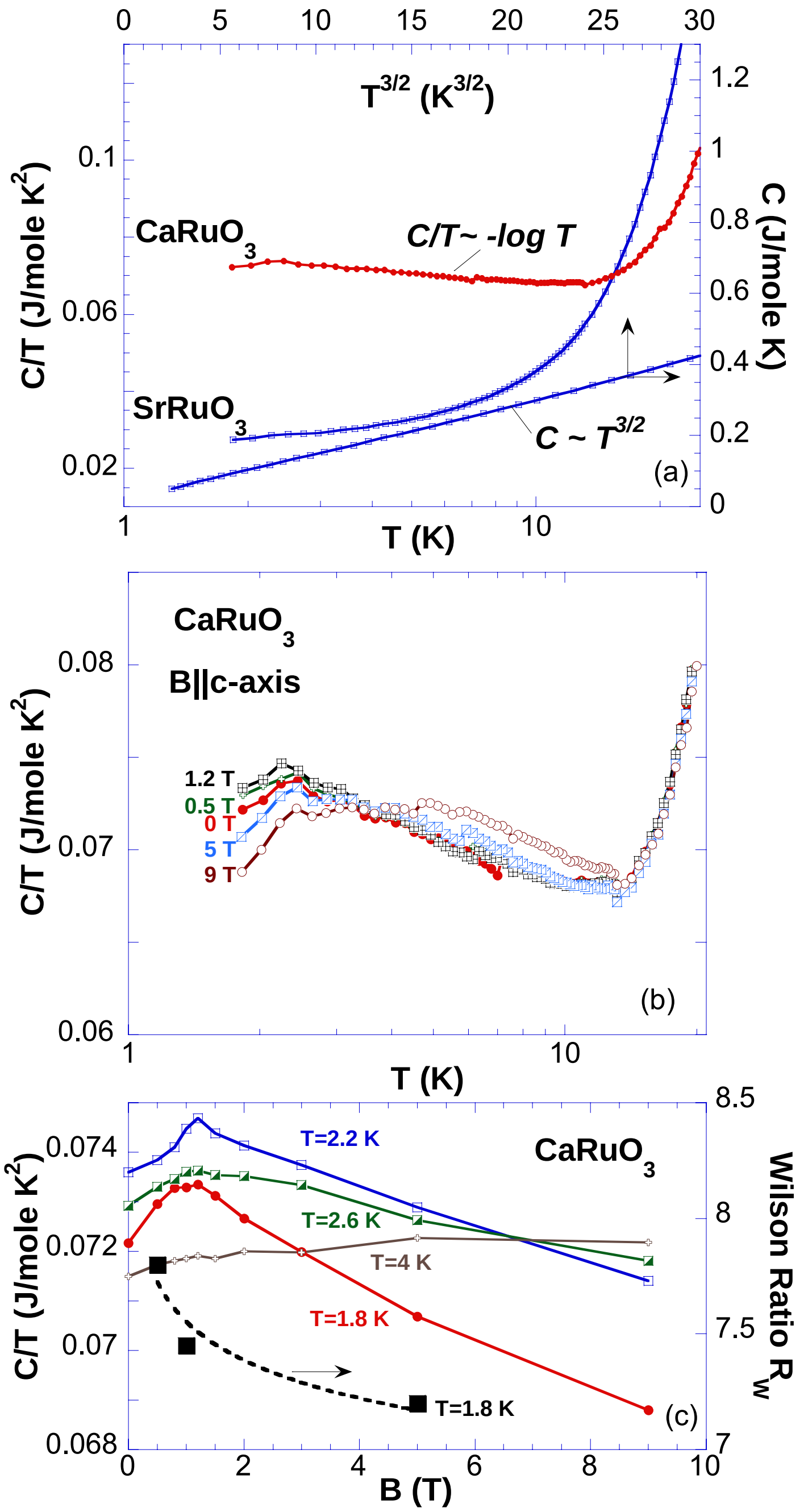

Fig. 3

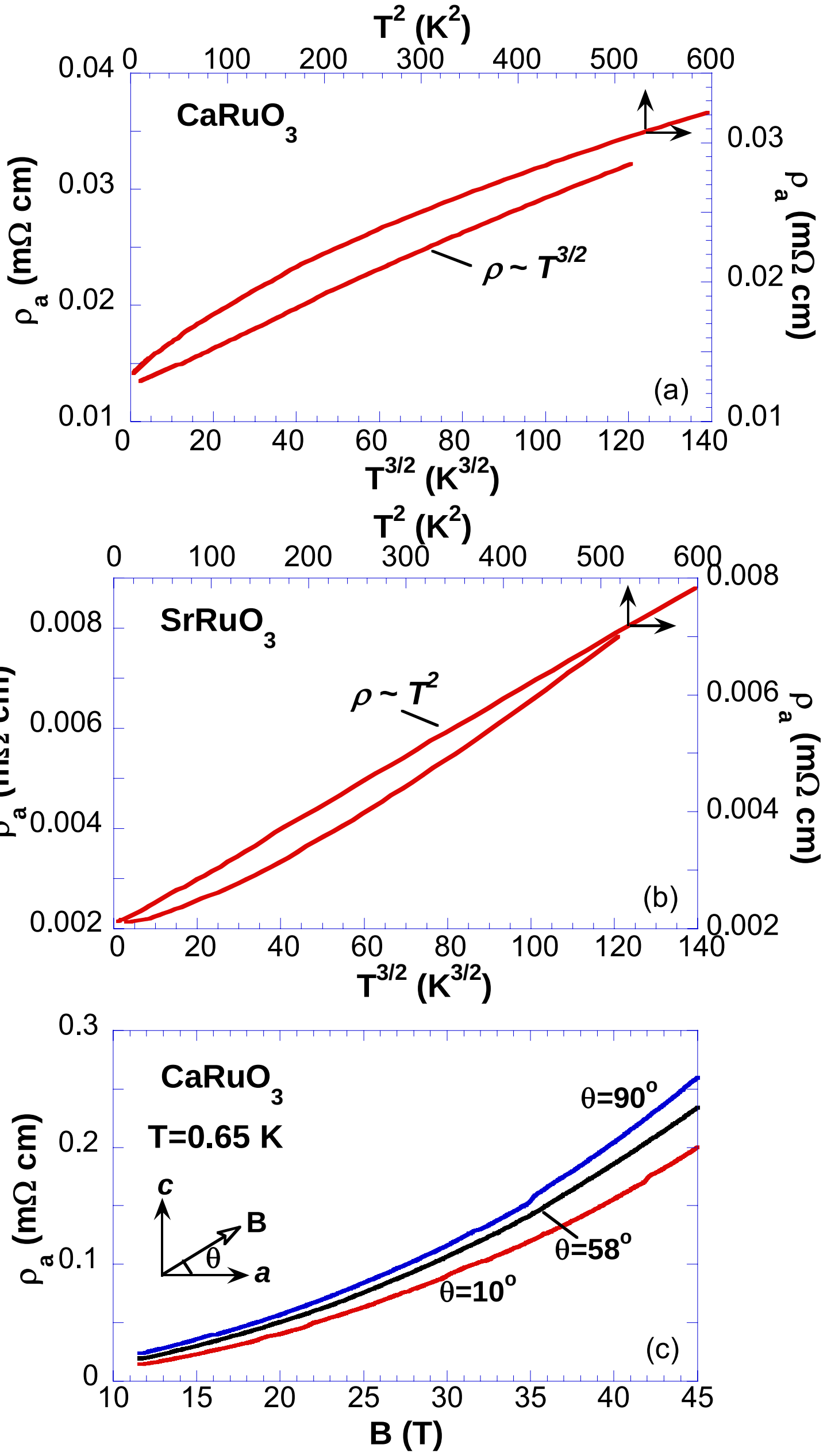

T² (K²)
0 100 200 300 400 500 600
CaRuO3
ρ ~ T^3/2
(a)
ρa (mΩ cm)
0.01 0.02 0.03 0.04
T^3/2 (K^3/2)
0 20 40 60 80 100 120 140
SrRuO3
ρ ~ T²
(b)
0.002 0.004 0.006 0.008
CaRuO3
T=0.65 K
c
B
θ
a
θ=90°
θ=58°
θ=10°
(c)
0 0.1 0.2 0.3
10 15 20 25 30 35 40 45
B (T)

Fig. 4